\documentclass[aps,prl,twocolumn,showpacs]{revtex4-1}

\usepackage[dvipdfmx]{color}
\usepackage[dvipdfmx]{graphicx}

\usepackage{dcolumn}% Align table columns on decimal point
\usepackage{bm}% bold math
\usepackage{multirow}
\begin{document}

\title{
A precaution for the hybrid density functional calculation of open-shell systems
}

\author{Jun-Ichi Iwata$^1$}
\author{Keisuke Sawada$^2$}
\author{Atsushi Oshiyama$^1$}

\affiliation{
$^1$Department of Applied Physics, The University of Tokyo, Tokyo 113-8656, Japan
 }
\affiliation{
$^2$Advanced Institute for Computational Science, RIKEN, Kobe 650-0047, Japan
 }
 \date{\today}

\begin{abstract}
We show that a naive treatment of open-shell systems in hybrid density functional calculations ignoring the spin dependence causes significant errors due to a kind of self interaction that is not emerged in spin-dependent calculations. As numerical examples, we compare the results of the LDA, GGA, and PBE0 calculations on the ionization potential and electron affinity of C$_{60}$ molecule and the GGA and HSE calculations on the singly charged monovacancy in crystalline Si.
\end{abstract}

\pacs{ } 

\maketitle

First-principles calculation based on the density functional theory (DFT) has been an indispensable tool for understanding, predicting, and designing materials properties \cite{hohenberg,kohn}. The usefulness is largely supported by the continuous development on the exchange-correlation (XC) functional. The hybrid-XC functional \cite{pbe0,hse03,hse06} becomes a new member of the workhorse functionals such as local-density approximation (LDA) \cite{pz81} and generalized-gradient approximation (GGA) \cite{pbe96}, and in many cases the results are better than those of GGA, especially for the band gap of semiconductors \cite{hsesol06}.

The hybrid-XC functional is constructed by a mixing of the LDA or GGA functional and the exact-exchange functional. The explicit form of the exact-exchange energy for spin-degenerate systems is written as \cite{hsesol06}
\begin{eqnarray}
E_x = -\frac{1}{2}\sum_{m,n} f_m f_n \int \!\!\! \int \!\! d{\bf r} d{\bf r}'
            \frac{ \phi_m^{*}({\bf r}) \phi_n^{*}({\bf r}') \phi_m({\bf r}') \phi_n({\bf r}) }{ |{\bf r} - {\bf r}' | },
\nonumber\\
\end{eqnarray}
where $\{\phi_n\}$ are the Kohn-Sham orbitals and $\{f_n \}$ are the corresponding occupation numbers which usually take $0$ or $2$ in spin-degenerate calculations. Spin-degenerate, or non-spin-dependent calculations are often performed in LDA or GGA even for open-shell systems when the spin-polarization effect is less important, and in that case the open-shell orbital is described by simply setting the occupation number as $f_n = 1$. Although the same approach is possible and indeed used \cite{vsihse} in the non-spin-dependent hybrid-XC calculations, however such naive treatment of open-shell orbitals causes significant errors due to a kind of self interaction that is caused by the exact-exchange term.

In this paper, we demonstrate the emergence of the self-interaction errors in non-spin-dependent hybrid-XC calculations for open-shell systems and the avoidance in spin-dependent calculations. As numerical examples, we compare the results of the LDA, GGA and the PBE0 hybrid functional \cite{pbe0} calculations on the ionization potential (IP) and electron affinity (EA) of C$_{60}$ molecule and the GGA and the HSE hybrid functional \cite{hse06} calculations on the formation energy of the singly charged monovacancy in crystalline Si.

 The calculations were performed with our real-space finite-difference pseudopotential code RSDFT \cite{iwata2} that has been developed for parallel computers including 10-PFLOPS-class systems \cite{hasegawa}. Troullier-Martins type norm-conserving pseudopotentials were used \cite{tm}, and the pseudopotentials generated with GGA were also used for hybrid-XC calculations.
  
  The first example is the IP and EA of C$_{60}$ molecule. The calculations were performed with the isolated boundary condition; the values of the wave functions were zero outside of the spherical simulation box of 20-bohr radius. With this boundary condition, charged state calculations can be performed without any artificial corrections, so that IP and EA are obtained from the direct difference between the total energies of each charge state, namely the $\Delta$SCF calculation.
The grid spacing was taken as 0.3 and 0.2 bohr for atomic structure optimization in LDA and GGA, respectively. We confirmed that the size of the simulation box and the grid spacings are large and fine enough to achieve convergence within sub meV. We also confirmed that the grid spacing of 0.4 bohr is enough for the calculations without atomic structure optimization.

The results of LDA, GGA, PBE0, and their spin-dependent versions (LSDA, spin-GGA, spin-PBE0, respectively) are summarized in Table \ref{c60}. For LDA and GGA, the structures are fully optimized in each charge state, and the resultant point-group symmetry is $I_h$ for the neutral system and $D_{5d}$ for the singly charged systems. We found that LDA and LSDA predict the experimental IP very accurately, but overestimate EA about 0.2 eV.  While GGA and spin-GGA predict experimental EA within 20 meV, but underestimate IP about 0.2 eV. We also found that the spin polarization effect is rather small: the total energy difference is 20 meV and 30 meV in the LDA and GGA, respectively, and the atomic structures are unchanged irrespective of the spin degree of freedom.

Next we compare the results of the (spin-)GGA and (spin-)PBE0 in Table \ref{c60}. The (spin-)PBE0 calculation was performed with the coarser grid spacing of 0.4 bohr and the fixed atomic structure obtained by the (spin-)GGA calculation. We confirmed that the (spin-)GGA values of IP and EA do not change with this coarser grid spacing and the fixed atomic structure.
As shown in Table \ref{c60}, PBE0 provides rather poor results comparing to GGA, while spin-PBE0 substantially improve the results; in particular the agreement with experimental IP is far better than (spin-)GGA.
\begin{table}
\begin{center}
\caption{\label{c60} 
LDA, GGA, and PBE0 results of the $\Delta$SCF calculations on the ionization potential (IP) and the electron affinity (EA) of C$_{60}$ molecule. The atomic structures are fully optimized for LDA and GGA in each charge state, and the atomic structures for PBE0 are the same as those of GGA.}
\begin{tabular}{ccc}
\hline
\hline
 & \ \ \ IP (eV) \ \ \ & \ \ \ EA (eV) \ \ \ \\
 \hline
LDA    \ \ \ &  \ \ \ 7.60 \ \ \  & \ \ \  2.92 \ \ \  \\
LSDA  \ \ \ &  \ \ \ 7.58 \ \ \  & \ \ \  2.94 \ \ \  \\
GGA   \ \ \ &  \ \ \ 7.37 \ \ \  &  \ \ \ 2.69 \ \ \  \\
spin-GGA  \ \ \ &  \ \ \ 7.33 \ \ \  &  \ \ \ 2.72 \ \ \  \\
PBE0   \ \ \ &  \ \ \ 7.87 \ \ \  &  \ \ \ 2.32 \ \ \  \\
spin-PBE0  \ \ \ &  \ \ \ 7.57 \ \ \  &  \ \ \ 2.59 \ \ \  \\
Expt.   \ \ \ & \ \ \  7.58 $\pm$ 0.04$^a$ \ \ \  & \ \ \  2.689 $\pm$ 0.008$^b$ \ \ \ \\
\hline
\hline
$^a$ Ref.~\cite{ipc60}\\
$^b$ Ref.~\cite{eac60}
\end{tabular}
\end{center}
\end{table}

\begin{figure}
\begin{center}
\includegraphics[width=0.7\linewidth] {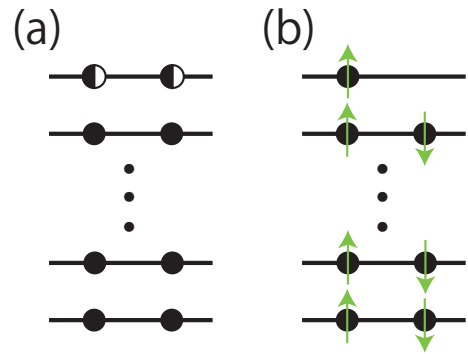}
\caption{\label{rhf}
(color online)
Schematic illustrations of the electron configurations with an open-shell orbital in spin-degenerate theory (a) and the spin-dependent theory (b). 
The half-shaded circle represents an electron with 50 \% of occupation.}
\end{center}
\end{figure}

The failure of the non-spin-dependent PBE0 calculations can be understood by comparing to the Hartree-Fock theory. The present PBE0 and spin-PBE0 calculation corresponds to the restricted Hartree-Fock (RHF) and the unrestricted Hartree-Fock (UHF) calculation, respectively. As well known in the quantum-chemistry community, RHF treatments of the open-shell systems require a special formulation, that is the restricted open-shell Hartree-Fock (ROHF) theory \cite{rohf}. However, in the present PBE0 calculations, we describe the open-shell orbital by just varying the occupation number, and the resultant electron configuration is the same as that of the ordinary RHF theory where each orbital is accommodated by two (half-charged) electrons as shown in Fig.~\ref{rhf}(a). As a consequence, one half-charged electron in the open-shell orbital feels the Coulomb potential originates from the other half-charged electron in the same orbital, and this is the source of the self-interaction error caused by the inadequate use of the RHF-like theory to the open-shell systems \cite{caution}.

In the spin-PBE0 calculations, the situation is similar to that of the UHF calculations. The exact-exchange energy in the UHF-like theory is written as
\begin{eqnarray}
&&E_x^{UHF} = \nonumber\\
&&-\frac{1}{2}\sum_{s}^{\uparrow,\downarrow}\sum_{m,n} f_{m,s} f_{n,s} \int \!\!\! \int \!\! d{\bf r} d{\bf r}'
            \frac{ \phi_{m,s}^{*}({\bf r}) \phi_{n,s}^{*}({\bf r}') \phi_{m,s}({\bf r}') \phi_{n,s}({\bf r}) }{ |{\bf r} - {\bf r}' | },
\nonumber\\
\label{euhf}
\end{eqnarray}
where the expression is just a sum of each spin term, and in this case the occupation numbers $\{ f_{n,s} \}$ take $0$ or $1$. The electron configuration for open-shell systems in this theory is schematically shown in Fig.~\ref{rhf}(b). Clearly the spin degree of freedom is adequately described in the UHF-like theory, and thereby no spurious interaction is emerged from Eq.~(\ref{euhf}). Thus the spin-PBE0 calculation can reveal its innate accuracy.
%
%%%%%%%%%%%%%%%%%%%%%%%%
%%%%%%%%%%%%%%%%%%%%%%%%
%%%%%%%%%%%%%%%%%%%%%%%%

The next example is the charged monovacancy in crystalline Si. DFT calculations for this system have a long history, and therefore many results have been obtained with various XC functionals and model sizes \cite{vsiao,vsinieminen,vsiog,vsipp,vsiwright,vsicm,vsihse}. Notwithstanding, the complete consensus seems not to be achieved yet. A HSE hybrid-XC calculation has been performed recently, but failed to reproduce the experimentally observed $C_{2v}$ structure for the singly-negatively charged monovacancy V$^{1-}$ \cite{vsihse}. 

We performed the HSE hybrid-XC calculations as well as GGA calculations for V$^0$ and V$^{1-}$ of Si monovacancies. The vacancy model is constructed from the 512-site supercell with optimized lattice constant of the perfect crystal: 5.438 and 5.465 {\AA} for HSE and GGA, respectively. The grid spacing was taken as 0.43 bohr, which corresponds to 52 Ry of the cut-off energy for plane-wave calculations. Structure optimizations were performed in each functional and charge state.

\begin{figure}
\begin{center}
\includegraphics[width=1.0\linewidth]{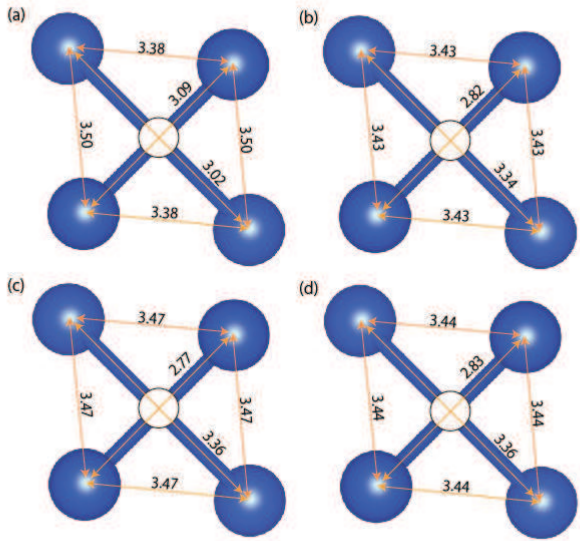}
\caption{\label{structure}
Inter-atomic distances around the vacancy of Si in singly-negatively charged state. The optimized structures are obtained from the HSE (a), GGA (b), spin-HSE (c), and spin-GGA (d) calculations with 512-site supercell. The unit is in \AA.}
\end{center}
\end{figure}

The optimized structure of the neutral vacancy is almost the same in both GGA and HSE. The symmetry is $D_{2d}$, and the two characteristic inter-atomic distances around the vacant site is 3.06 and 3.53 {\AA} for GGA, and 2.98 and 3.54 {\AA} for HSE. In Fig.~\ref{structure}, we show the optimized structure of V$^{1-}$ obtained by the HSE, GGA, and their spin-dependent version (spin-HSE and spin-GGA, respectively) calculations. Both GGA and spin-GGA provide essentially the same structure of $C_{2v}$ symmetry that is consistent with the ENDOR measurement \cite{endor}. Spin-HSE also provides the $C_{2v}$ structure (Fig.~\ref{structure}(c)) for V$^{1-}$, but only HSE provides a different structure of $C_2$ (approximately $D_2$ \cite{comment}) symmetry (Fig.~\ref{structure}(a)). The previous HSE calculation \cite{vsihse} also reported the $D_2$ structure for V$^{1-}$ .

The formation energy of the monovacancy is defined as
\begin{eqnarray}
E_f = E^{V^q}_{N-1} + \mu_{Si} + q(\varepsilon_F + \varepsilon_v) - E^{host}_N,
%\nonumber
\end{eqnarray}
where the total energy of the monovacancy in $q$ charge state, chemical potential of the bulk Si, Fermi energy relative to the valence-band top of the bulk Si, energy of the valence-band top, and the total energy of the host Si crystal of $N$-atom supercell are appeared in the right-hand side of the equation. The corrections for charged state calculations are according to Ref.~\cite{lany}. The formation energy obtained by the (spin-)HSE and (spin-)GGA calculations are summarized in Table~\ref{eform}. The difference between the GGA and spin-GGA results indicates the spin-polarization effect is about 90 meV in the formation energy of V$^{1-}$. However, the formation energy difference between HSE and spin-HSE is much larger 240 meV. Taking into account the inconsistent optimized structure and the large energy difference, we conclude that the non-spin-dependent HSE calculations for singly charged Si monovacancies also suffer from the self-interaction error caused by the exact-exchange term.
\begin{table}
\begin{center}
\caption{\label{eform} 
HSE and GGA results of the formation energy of Si monovacancy in neutral and singly-negatively charged state with $\varepsilon_F=0$. The unit is in eV.}
%\begin{tabular}{ccccccc}
%\hline 
%\hline
%     & \ \ \ GGA &  \ \ \ spin-GGA  &  \ \ \ HSE &  \ \ \ spin-HSE &  \ \ \ Expt.   \\
%\hline
%V$^0$  & \ \ \ 3.61 & \ \ \ - & \ \ \ 4.19 & \ \ \ - & \ \ \ 2.1 - 3.6 & \\
%V$^{1-}$ & \ \ \ 4.13 & \ \ \ 4.04 & \ \ \ 5.12 & \ \ \ 4.88 & \ \ \   &  \\
%\hline
%\hline
\begin{tabular}{ccc}
\hline
\hline
                  & \ \ \ V$^0$ \ \ \ & \ \ \ V$^{1-}$ \ \ \ \\
\hline
 HSE          & \ \ \ 4.19 \ \ \    & \ \ \ 5.12 \ \ \ \\
 spin-HSE  & \ \ \ 4.19 \ \ \    & \ \ \ 4.88 \ \ \ \\
 GGA         & \ \ \ 3.61 \ \ \    & \ \ \ 4.13 \ \ \ \\
 spin-GGA & \ \ \ 3.61 \ \ \    & \ \ \ 4.04 \ \ \ \\
 Expt.         & \ \ \ 3.6 $\pm$ 0.5$^a$ \ \ \ & \\
                  & \ \ \ 3.85 $\pm$ 0.15$^b$ \ \ \ & \\
 \hline
 \hline
 $^a$ Ref.~\cite{watkins}\\
 $^b$ Ref.~\cite{suezawa}
\end{tabular}
\end{center}
\end{table}

In conclusion, we have demonstrated that the naive treatment of the open-shell orbital in spin-degenerate hybrid-XC calculations suffer from the self-interaction error that is not emerged in spin-dependent calculations.  As numerical examples, we have performed the calculations on the IP and EA of C$_{60}$ molecule and the singly charged monovacancy in crystalline Si. Even in the case that the spin-polarization effect is considered to be less important, we should perform the spin-dependent calculations when we apply the hybrid-XC functional for open-shell systems to avoid the artificial error that doesn't exist in the intrinsic theory.

\begin{acknowledgments}
This work was supported by "Computational Materials Science Initiative", conducted by MEXT, Japan.  Computations were performed mainly at K Computer in Advanced Institute for Computational Science, RIKEN and at Supercomputer Center in ISSP, University of Tokyo, and CCS, University of Tsukuba.
\end{acknowledgments}


\begin{thebibliography}{99}
\bibitem{caution}
Since the ROHF-like implementation of the hybrid-XC functional is not standard in the first-principles calculation programs, the users of these programs must be careful by themselves to avoid the errors described here.
\bibitem{comment}
In Fig.~\ref{structure}(a), if we identify the length 3.02 and 3.09 {\AA}, the structure becomes $D_2$.
\end{thebibliography}
\end{document}